\documentclass[reprint]{JASA}
\renewcommand{\today}{17 November 2026}
\begin{document}
\title[Music of the Deep]{Music of the Deep: An acoustic blueprint of sperm whale codas}

\author{Mudit Sinha}
\email{muditsinha01@gmail.com}
\author{Sanika Chavan}
\email{sanikac10@gmail.com}
\affiliation{Independent Researcher, San Bruno, California 94066, United States}

\preprint{Manuscript prepared for the Journal of the Acoustical Society of America}
\date{17 November 2026}

\begin{abstract}

Sperm-whale codas are usually described by click count and spacing. Earlier work found additional structure in click waveforms and across coda sequences. We ask how these two levels compare with performed music. We examine 1,952 codas alongside violin performances, drum grooves, and sung melodies. Using frozen audio encoders, we describe constituent sounds and complete units separately, comparing composition, order, and sequence context. In codas, click composition improves prediction of whole-coda acoustic form beyond timing. Click-form order carries additional conditional information, and rearranging click spacing changes the acoustic representations across six encoders. Knowing the second preceding coda improves held-out prediction beyond knowing only the last one. The musical references show different combinations of these relationships, so similarities across sequences do not imply similar organization within units. We then examine acoustic properties that sequence measurements cannot capture. Codas span broad, reproducible ranges along dimensions validated using human emotion ratings. WhAM-generated clips retain measurable acoustic differences from matched natural codas. Together, these findings show why a coda cannot be described by timing, constituents, or sequence context alone. They motivate an acoustic blueprint that brings these measurements together and provides a reference for comparing natural and generated sound.

\end{abstract}
\maketitle

\section{Introduction}

Sperm-whale codas are short sequences of clicks produced during social interactions. Studies have traditionally described them by their click count and the time between clicks~\cite{watkins1977,pavan2000,weilgart1997,rendell2003,rendell2005,gero2016}. These measurements reveal differences among individuals, social groups, and calling contexts~\cite{gero2016,sharma2024}, but they do not describe all the variation in recorded click waveforms or the relationships between successive codas.

Earlier work in \emph{Rhythm of the Deep}~\citep{sinha2026} (hereafter \emph{Rhythm}) found differences in click sound beyond timing and predictive relationships across successive codas. These findings concern two levels of organization. One describes how smaller sounds form a coda; the other describes how complete codas follow one another. Music offers a useful comparison because notes and percussion events also form larger units, and earlier events can shape expectations about what follows~\cite{patel2003,pearce2012}. We ask whether the measured relationships between these levels resemble those found in performed music.

\begin{figure*}[!t]
\centering
\includegraphics[width=0.98\textwidth]{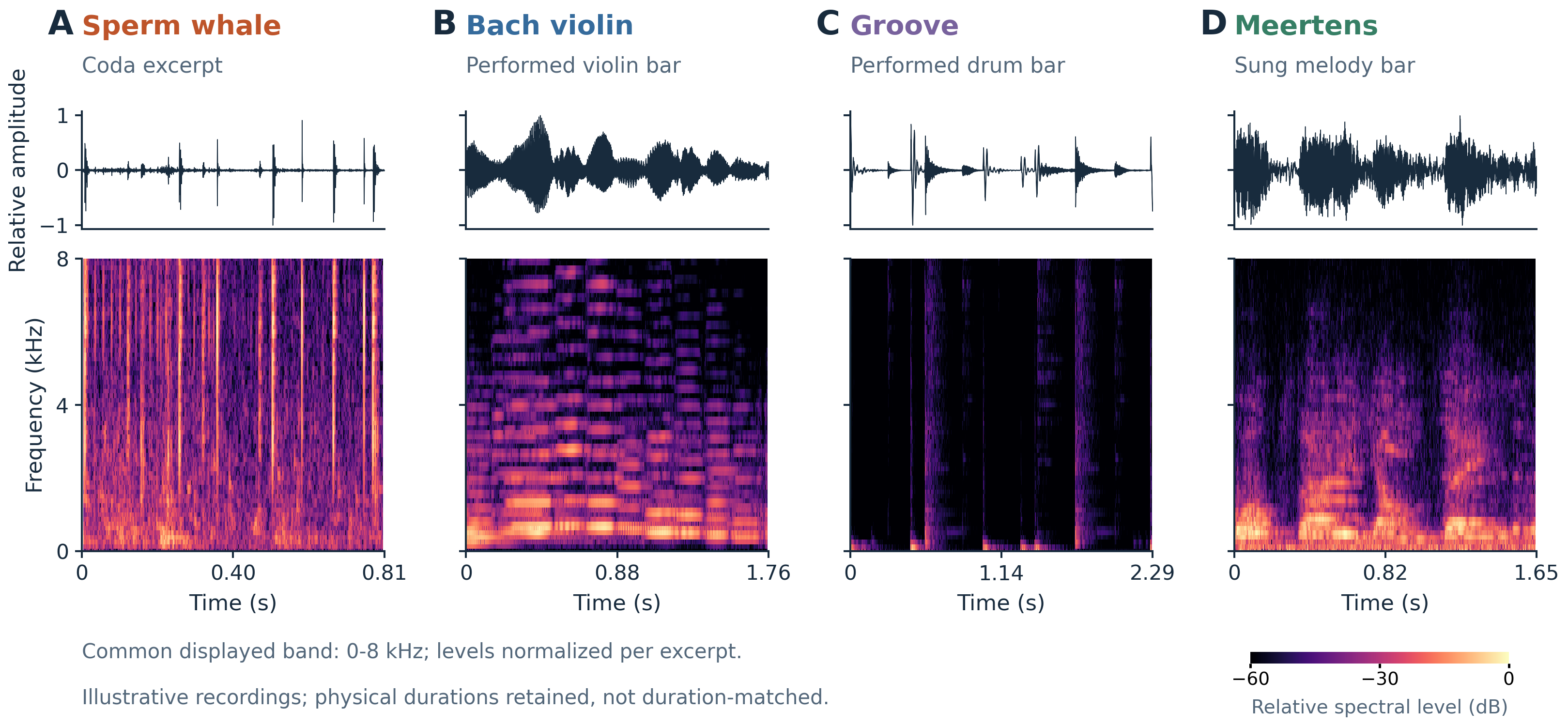}
\caption{A CETI coda and three performed musical excerpts, shown as recorded waveforms and spectrograms at their original durations.\label{fig:examples}}
\end{figure*}

We compare codas with performed Bach violin music, drum grooves, and sung melodies (Fig.~\ref{fig:examples}). We first measure how often complete units repeat and how much the current unit tells us about the next one. We then test how constituent sounds contribute to each unit and whether their order adds information beyond their composition. For codas, we also rearrange click spacing while retaining the recorded click cores. Finally, we ask whether earlier complete units improve prediction of what follows. These measurements let us compare the two levels together rather than infer one from the other.

A whale click is not a musical note, and the recordings need not have equal acoustic repertoires. We use frozen audio encoders to describe similarities among waveforms within each corpus. We then ask the same questions about the resulting acoustic groups, while retaining each corpus's own event definitions and recording boundaries. Holding out complete sources tests whether the measured relationships extend beyond the recordings used to form those groups. The common procedure makes the comparisons possible without assigning the same meaning to whale and musical units.

The comparison of sequences still leaves another question about the recorded sounds. Music varies acoustically along dimensions associated with human ratings of emotion~\cite{juslin2003}. We test whether codas show similarly broad and reproducible variation along these dimensions. This adds an acoustic measurement that neither constituent ordering nor prediction of the next unit can provide.

Our results show that the comparison depends on which level is measured. The musical references combine constituent composition, order, and earlier sequence context differently from codas, even though all four corpora have measurable sequence regularity. Codas also span a broad range along human-validated acoustic reference directions. These findings motivate an \emph{acoustic blueprint} that describes each property separately and brings them together into a fuller account of the recorded sound.

The blueprint also provides a way to examine generated audio. We compare isolated clips from WhAM~\cite{paradise2025} with natural recordings to see which measured properties differ. We first describe all available generated clips and then compare the subset that can be balanced on measured waveform properties. That smaller comparison asks a stricter question, although it cannot establish equivalence of the original recording equipment or geometry.

\section{Related Work}

Studies of coda repertoires established recurring click-count and timing patterns across callers and social groups~\cite{watkins1977,rendell2003,rendell2005,gero2016}. Acoustic studies also measured click spectra and envelopes, showing properties that click timing alone cannot describe~\cite{mohl2003,begus2025}. \emph{Rhythm} investigated click waveforms and whole-coda sequences together~\cite{sinha2026}. We extend that question by testing the relationship between these levels against musical recordings.

Music offers examples of constituent sounds forming larger units and of earlier units shaping what follows~\cite{patel2003,pearce2012}. Performed timing varies around an established pattern~\cite{repp1992,friberg1999}. Sharma and colleagues found gradual duration changes across adjacent codas within a shared rhythm and tempo type~\cite{sharma2024}. Those findings provide background for our comparison, but the present analysis tests constituent ordering and earlier whole-coda context rather than caller-specific timing.

Pretrained audio encoders make it possible to describe recordings without assuming that a whale click and a musical note are equivalent~\cite{hagiwara2023,miron2026,hsu2021,baevski2020,garcia2023,openbeats2025}. Such comparisons must still account for recording differences. Source orientation, propagation, and equipment can change the measured spectra~\cite{mohl2000,schulz2009,thode2002,zimmertyack2005}. We therefore evaluate acoustic groups on unseen recordings and retain direct acoustic and recording-variable controls.

Individual acoustic cues have been reported in sperm-whale codas~\cite{antunes2011,oliveira2016}. Human-rated music and environmental sounds also provide reference directions for valence and arousal~\cite{russell1980,aljanaki2017,fan2017}. We test variation along those independently validated acoustic directions without assigning emotions to whales. Finally, WhAM has demonstrated the generation of isolated coda-like audio~\cite{paradise2025}. We ask whether those clips also match natural recordings across the frozen encoder representations, rather than judging their audible resemblance alone.

\begin{table*}[!t]
\caption{Questions, measurements, and controls used for the CETI comparison.\label{tab:methodmap}}
\centering\footnotesize
\begin{tabular}{lll}
\parbox[t]{0.24\textwidth}{\raggedright Question} & \parbox[t]{0.34\textwidth}{\raggedright Measurement} & \parbox[t]{0.30\textwidth}{\raggedright Main comparison or control} \\
\hline
\parbox[t]{0.24\textwidth}{\raggedright How do coda sequences compare with music?} & \parbox[t]{0.34\textwidth}{\raggedright Repetition, successor variety, and prediction from preceding units} & \parbox[t]{0.30\textwidth}{\raggedright Held-out sources, native sequence boundaries, frequency-matched references, and alternative predictors} \\
\parbox[t]{0.24\textwidth}{\raggedright How do smaller sounds contribute to each unit?} & \parbox[t]{0.34\textwidth}{\raggedright Composition beyond timing; ordered versus bagged constituents; conditional information} & \parbox[t]{0.30\textwidth}{\raggedright Exact constituent bags, held-out source tests, and separate first-order-model controls} \\
\parbox[t]{0.24\textwidth}{\raggedright Are learned coda groups acoustically grounded?} & \parbox[t]{0.34\textwidth}{\raggedright Extra state-prediction accuracy from conventional spectral descriptors} & \parbox[t]{0.30\textwidth}{\raggedright Source-held-out nuisance-only model and common-band sensitivity} \\
\parbox[t]{0.24\textwidth}{\raggedright How broad is emotion-related acoustic variation?} & \parbox[t]{0.34\textwidth}{\raggedright Reliable rank positions and spread along validated human-reference directions} & \parbox[t]{0.30\textwidth}{\raggedright Human eligibility, recording-relative level adjustment, and supported structural sensitivity} \\
\parbox[t]{0.24\textwidth}{\raggedright How do generated clips compare with natural codas?} & \parbox[t]{0.34\textwidth}{\raggedright Six-encoder natural--synthetic acoustic-distribution gaps} & \parbox[t]{0.30\textwidth}{\raggedright Disjoint duration-matched donor panels and a separate acquisition-balance feasibility audit} \\
\hline
\end{tabular}
\end{table*}

\newpage
\section{Methods}

We compare codas with music, first as sequences of complete sounds and then through the smaller sounds within each unit. We next measure whether the acoustic groups correspond to conventional spectral properties and examine variation along human emotion-reference directions. Finally, we compare natural codas with isolated WhAM recordings. Table~\ref{tab:methodmap} links each question with its measurement and control.

\subsection{Recordings and analysis sets}

We use 63 sperm-whale recordings from the publicly shared \emph{SpermWhales} collection linked in reference 34 of Gubnitsky and Diamant~\cite{gubnitsky2024}. They include 33 remote WAV recordings and 30 tag FLAC recordings, each with matching annotations. From these recordings, we selected 1,952 codas containing 10,755 annotated clicks for analysis. The Supplement identifies the source files and their checksums. Gubnitsky et al.~\cite{gubnitsky2025} describe related methods for automatic coda detection and annotation. The final companion \emph{Rhythm} study uses the same 63-recording collection, but asks different questions about click and whole-coda organization. We use the supplied waveform intervals and source identifiers, and do not assign caller, date, social-unit, or coda-type labels that have not been verified for these recordings. The exported files contain 60- and 96-kHz recordings, but the available headers do not establish the original hydrophone, animal identity, or recording geometry. The Supplement gives the release checks and analysis-specific counts.

The upper-tier analysis uses all 1,952 codas. To examine click constituents without shared crop material or overlapping click windows, the strict lower tier retains 1,132 codas and 6,039 clicks. Earlier-coda histories remain within 195 ordered annotation segments. Four conservative recording/source pools define the held-out blocks; they are not four independently identified animals. The 182 groups formed by an eight-second-gap rule are operational groups, not confirmed continuous behavioral bouts. We do not treat ordering within an annotation segment as a sequence from one known caller.

The musical references include six performed Bach violin works, Groove drum recordings grouped into 20 performer/session sources~\cite{gillick2019}, and Meertens sung melodies grouped into nine singer/tune components~\cite{vanKranenburg2014}. The Meertens analysis links symbolic MTC-FS-1.0 transcriptions to their original MTC-OGLAUDIO-1.0 field recordings. Selected bars are aligned to pitch-verified melodic occurrences; numbered lyric-verse identity was not independently verified. We retain the recorded notes or drum events and each source's own bar definitions. Source-held-out tests keep the original gaps between selected bars rather than joining unrelated passages.

\subsection{Frozen acoustic representations}

Before examining organization, we need a consistent way to compare recorded sounds. Six frozen encoder families provide waveform representations, using HuBERT, wav2vec~2.0, AVES, AVES2, OpenBEATs, and VampNet~\cite{hsu2021,baevski2020,hagiwara2023,miron2026,openbeats2025,garcia2023}. The analysis uses three prespecified layers from each of the first five families and one VampNet representation. Within the discovery recordings, we standardize the representations, reduce their dimensions, and group acoustically similar waveforms. Agreement balanced across encoder families defines the primary acoustic states. We repeat these fitting steps without each held-out source. The resulting groups are measured acoustic labels, not established biological coda types.

We fix 32 states for the primary comparison and use 16 and 64 states to examine sensitivity to grouping resolution. Each corpus can use those groups at different frequencies. We therefore report effective state occupancy alongside repetition and prediction uncertainty. The same procedure does not make a musical bar and a coda equivalent units.

\subsection{Whole-unit continuation and prediction}

We first ask how complete units follow one another. Adjacent repetition measures how often a state follows itself. Conditional entropy measures uncertainty about the next state after the current one is known. Dividing by the corpus's own state entropy puts this uncertainty in relation to the acoustic variety measured in that corpus. We also count the variety of possible successors and compare it with references preserving the state frequencies. Predictions use native sequence segments and do not cross recorded gaps.

We then compare prediction from one, two, and three preceding states at common eligible positions. The primary order-0--3 ladder uses 1,478 CETI targets with three predecessors; its strict cohort has 374 targets. A separate one-versus-two-predecessor comparison uses 1,605 upper-tier targets, or 528 on the strict cohort. We do not exchange these denominators. Run-collapsed sequences and repetition-aware models ask different questions from the pure order comparison. The Supplement reports the choice of model, state resolution, matching, and source uncertainty for each test.

We also ask whether the observed prediction gain depends on more than the recorded frequencies of adjacent state pairs. For this check, we rearrange the state sequences within each original chronology segment while preserving every adjacent-pair count, the segment length, and its endpoints. The fitted predictors and the same 1,478 target positions remain fixed. We compare the real gain with 1,999 such sequence rearrangements and correct across the 21 encoder-readout and state-resolution tests. Unlike a fitted first-order generator, this reference retains the exact observed neighboring-state counts. It tests additional conditional sequence information within these recordings, not generalization to new biological sources.

\subsection{Constituent composition and order}

We next ask what the smaller sounds contribute to each complete unit. The coda constituents are verified click windows from the strict lower-tier cohort; the musical constituents are the original notes or drum events. We first predict the unit's acoustic form using its count, duration, span, and timing measurements. We then add the types of constituent sounds present. The held-out accuracy gain measures what composition contributes beyond the simpler description.

Composition does not specify which sound occupies each position. We therefore also compare the real ordered constituents with rearrangements that keep the same constituents, their counts, and their original time slots. This exact-bag control asks whether assigning these sounds to their observed positions adds predictive information. A separate information test compares the conditional structure against sequences generated by a fitted first-order model. It does not preserve every observed transition count and is not interchangeable with the held-out order-2 predictive gain. The planned comparisons and correction families are reported in the Supplement~\cite{holm1979}.

We also test how the spacing of clicks contributes when the click sounds themselves are kept. For each of the 1,132 strict-cohort codas, we rearrange its own annotated inter-click intervals. The click count, set of intervals, total duration, and first and last click positions stay fixed. We preserve short 6-ms waveform cores at the new click locations and reconstruct the intervening background. A separate roundtrip-resampling sham processes the background while leaving the original click times unchanged. All six frozen encoder families measure the difference from the original audio. This comparison tests sensitivity to a change in timing, although the sham cannot exactly match every effect of resampling.

\subsection{Spectral grounding and emotion-reference directions}

An encoder can separate sounds without telling us which measured acoustic features distinguish them. We therefore test whether conventional spectral descriptors improve the prediction of whole-coda acoustic states beyond recording-related variables. The primary test uses native-band measurements, including the ratio of high- to low-frequency energy, and reports ordinary and balanced accuracy on the same held-out predictions. Source pools are excluded from every correction and fitting step. We also examine supported recording-variable adjustments and a common 16-kHz processing band. These tests ask whether conventional spectral measurements provide additional evidence for the learned groups under the available recording controls.

We then ask whether codas show a broad, reproducible range along acoustic directions associated with human ratings of emotion. We learn valence and arousal directions from DEAM music and EmoSoundscapes audio~\cite{russell1980,aljanaki2017,fan2017}. The directions must predict held-out human ratings before we apply them to codas. Of the 24 combinations of six encoders, two references, and two axes, 23 pass that check. We test whether codas retain reliable relative positions and whether their central 80\% spread falls within a prespecified factor of two of the corresponding human-audio spread.

We first examine the original projections and then adjust sound level relative to each recording. The correction is fitted without the held-out source. A stronger sensitivity also adjusts for duration, click count, and typical inter-click interval where the data support that comparison. It includes 1,839 codas and evaluates the other conditions on those same rows; it does not replace the full-cohort result. Bootstrap intervals preserve source and human-reference groups. These measurements test variation along human-validated acoustic directions, not what whales feel.

\subsection{Generated codas and acquisition controls}

We first compare all 45 isolated WhAM clips with disjoint natural panels matched for duration~\cite{paradise2025,gretton2012}. We measure distribution differences between encoder representations using squared maximum mean discrepancy (MMD$^2$)~\cite{gretton2012}. Subtracting the natural-to-natural difference from the generated-to-natural difference gives the excess MMD$^2$ gap. Repeating donor selection 999 times gives a descriptive range for the fixed clips. This range measures panel sensitivity, not uncertainty over whale populations or future generated recordings.

We next ask whether the comparison survives stronger matching on the measured waveforms. Both sets are processed at 16 kHz with the same 300--7500-Hz bandpass and RMS level. We match log duration, an energy-based noise proxy, crest at fixed level, spectral centroid, spectral bandwidth, and raw duration. Only 12 of the 45 generated clips lie within every corresponding natural marginal range. We therefore compare these 12 clips with two disjoint panels of 12 natural codas each. The largest standardized difference across the generated and natural panels is 0.156, below the declared balance limit of 0.20.

The original strict 0.5-standard-deviation pair caliper could not yield a feasible matched sample. A documented maximum-sample sensitivity widened this caliper to 1.5 while retaining the same balance threshold. Matching used input measurements, not the encoder gaps, and the acoustic test followed after the panels were fixed. We report both ordinary conditional 95\% reweighting intervals and simultaneous intervals across the six encoder representations. These intervals describe the supported finite panels. Common digital processing cannot make the original hydrophones or recording geometries equivalent, and the 33 excluded generated clips are outside the scope of this balanced comparison.

\section{Results}

\begin{figure*}[t]
\centering
\includegraphics[width=0.98\textwidth]{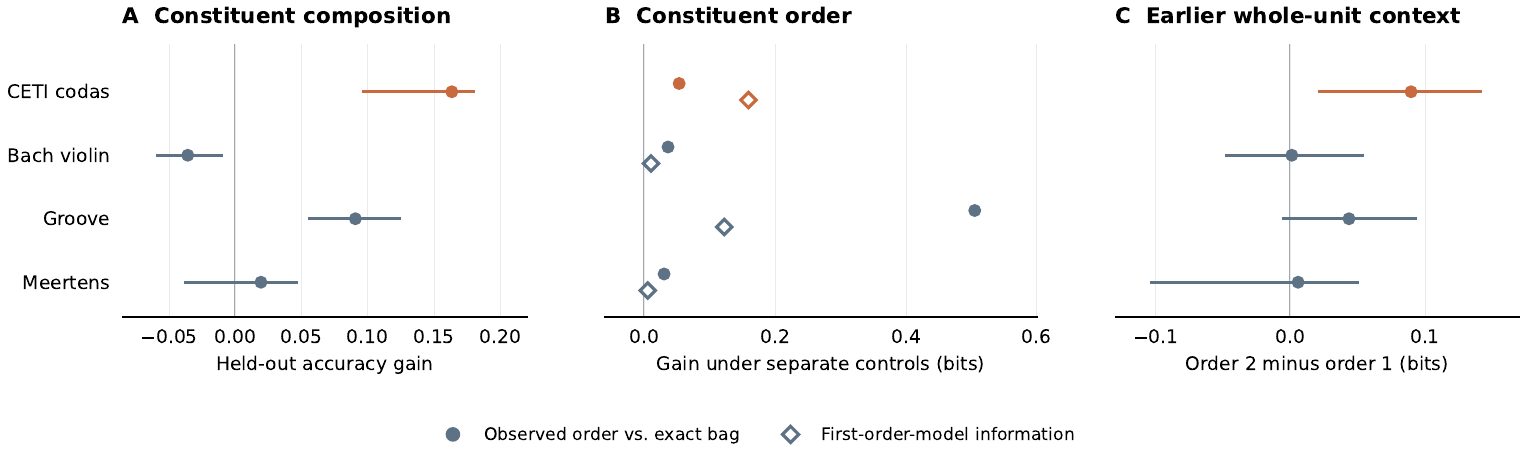}
\caption{Three parts of the acoustic comparison at $K=32$. (A) Composition gain beyond timing. (B) Separate constituent-order tests using exact-bag (circles) and fitted first-order (diamonds) references. (C) Prediction gain from a second preceding whole unit. Intervals in A and C condition on fitted models.\label{fig:profile}}
\end{figure*}

\subsection{Coda sequences and the musical references use different acoustic profiles}

All four corpora have measurable sequence regularity, but uncertainty and repetition take different forms. At the primary 32-state resolution, normalized uncertainty about the next unit is 0.805 for codas, 0.515 for Groove, 0.590 for Bach violin, and 0.789 for Meertens. Codas repeat the same state in 44.1\% of adjacent pairs, compared with 41.7\% for Groove, 40.9\% for Bach violin, and 18.9\% for Meertens. Similarity in one measure therefore does not imply similarity in the other.

The same 32-state representation also gives different effective occupancies. Codas use 6.44 effective states, against 11.63 for Bach violin, 8.14 for Groove, and 16.69 for Meertens. These occupancies help explain why repetition and normalized uncertainty should be read together rather than interpreted as equivalent acoustic organizations. Successor counts and frequency-matched controls appear in the Supplement. Within codas, finer state resolutions also produce more unseen held-out transitions, as expected when smaller groups create more possible distinctions.

\subsection{Constituent composition and order make different contributions}

Knowing which click forms occur improves prediction of a coda's acoustic form beyond timing alone. At 32 states, adding composition increases held-out accuracy by 0.163 for codas, compared with 0.091 for Groove, 0.019 for Meertens, and $-0.036$ for Bach violin. The direct difference in gain is positive relative to Bach violin and Meertens after correction across the three musical references; the corrected interval includes zero for Groove. The same measurement therefore finds different contributions from constituent composition in the tested recordings (Fig.~\ref{fig:profile}).

Click-form order contributes information under the separate first-order reference. The excess is $0.159$ bits for codas (Holm-adjusted $p=0.014$). Groove also gives a corrected positive result, whereas Bach violin and Meertens do not. The held-out exact-bag test asks a narrower question. It compares the recorded constituent order with rearrangements of those same sounds at the same time slots. Its coda gain is $0.053$ bits per target, but the raw $p=0.0185$ becomes $p=0.574$ after correction. Thus, conditional order information is detected, while an additional held-out order-2 advantage is not established by that different test (Fig.~\ref{fig:profile}).

\subsection{Changing click timing changes the measured sound}

Click timing matters even when the recorded click cores are retained. Rearranging the original inter-click intervals moves the encoder representations farther from the recordings than the background-resampling sham does. The mean paired excess standardized distance ranges from 0.608 to 0.872 across the six frozen encoder families. All six conditional intervals, including their simultaneous six-encoder adjustments, are positive. The effect therefore appears across different representations of the waveforms, rather than depending on one encoder.

The exact one-sided test across the four source pools gives $p=0.0625$ for each family, so the result is not independently significant at that level. The resampling sham also cannot reproduce every waveform change made by the timing edit. The supported finding is therefore sensitivity to the defined intervention, not a clean estimate of timing's effect in isolation. Alongside the composition analysis, it shows that the sounds present and their placement in time capture different acoustic information.

\subsection{Earlier whole codas provide conditional predictive context}

Knowing two preceding codas instead of one improves prediction in the primary held-out model. At 32 states, the gain is $0.090$ bits per target, with a conditional source-bootstrap interval of $[0.020,0.143]$. The exact test across four conservative source pools gives $p=0.0625$, so the gain is not independently significant at the source level. Under the same primary view, the gains are 0.001 for Bach violin, 0.044 for Groove, and 0.006 for Meertens.

A separate first-order-model information control finds $0.170$ excess bits among complete codas (Holm-adjusted $p=0.014$). That result is consistent with additional conditional structure, but it is not the same estimate as the held-out predictive gain.

The adjacent-pair-preserving rearrangement provides a stricter check on this sequence result. It keeps every recorded neighboring-state count within its original segment. The observed $0.090$-bit gain is $0.0156$ bits per target above that reference, with conditional $p=0.008$ before and $p=0.080$ after correction across 21 tests. The separate fitted first-order control and this exact-pair reference therefore measure related but different departures from neighboring-state patterns. Neither changes the four-pool source uncertainty. Repetition-aware models and the 16- and 64-state sensitivities also change the prediction question rather than provide interchangeable estimates (Fig.~\ref{fig:profile}).

\subsection{Measured spectra provide a limited additional acoustic check}

Conventional spectral descriptors do not provide a reliable additional held-out prediction gain for the learned coda states. At the primary $K=32$ resolution, native-band descriptors increase ordinary accuracy by only 0.0031, or 0.31 percentage points (exact four-source $p=0.4375$). Balanced accuracy does not improve in the same test.

A within-recording assignment control shows conditional correspondence between spectral features and learned states, but its ordinary-accuracy result does not survive the planned 18-test correction ($p=0.085$). Adjusting for supported recording variables or using a common processing band changes the estimates without establishing a four-source gain. The encoder groups therefore show some measurable spectral relationships, but these conventional descriptors do not independently predict the groups across held-out CETI sources. The full comparisons appear in the Supplement.

\subsection{Codas span broad emotion-related acoustic ranges}

Codas occupy a broad, reproducible range along acoustic directions established using human emotion ratings. Of 24 encoder, reference-dataset, and axis combinations, 23 pass validation on human audio. The excluded combination is wav2vec~2.0 with EmoSoundscapes valence. Coda rank positions are reliable in all 23 eligible comparisons under the prespecified test, with adjusted empirical $p=0.001$, the minimum attainable from 999 permutations.

\begin{figure*}[t]
\centering
\includegraphics[width=0.98\textwidth]{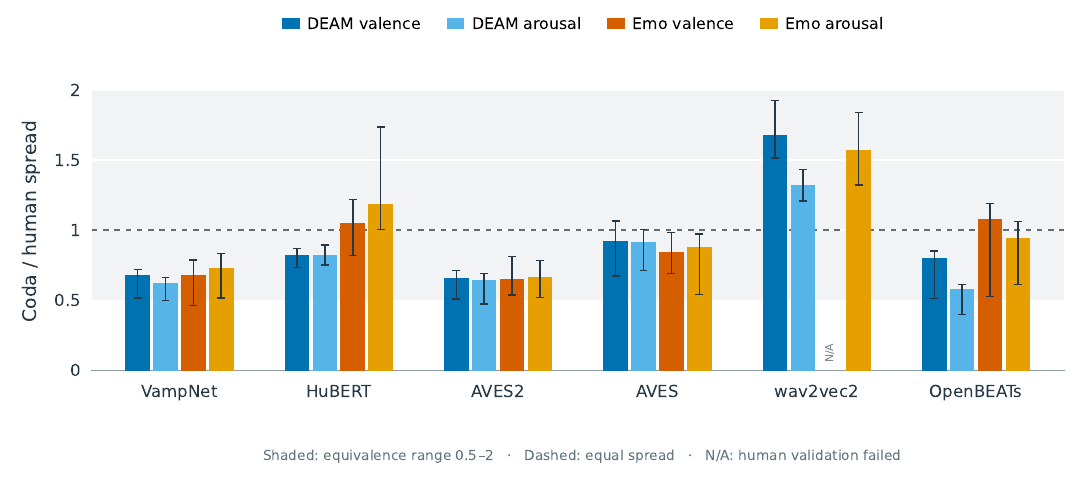}
\caption{Coda-to-human central-80\%-spread ratios after recording-relative level adjustment. Bars show six encoders across valence and arousal references with simultaneous intervals. Shading marks the 0.5--2.0 equivalence bounds; N/A marks failed human-reference validation.\label{fig:affect}}
\end{figure*}

The central acoustic spread lies within the prespecified factor-of-two human-reference bounds in 20 of 23 eligible raw comparisons. The same 20 of 23 remain within those bounds after adjusting sound level relative to the recording, using all 1,952 codas. A stronger adjustment for duration, click count, and ICI is supported for 1,839 codas and retains 11 of 23. The broad range is therefore reproducible under the primary level adjustment, while the stronger structural control changes how many comparisons meet the bounds. These are different adjustments of related data, not independent replications or measurements of whale emotions (Fig.~\ref{fig:affect}).

\subsection{Generated clips retain measured gaps on the supported comparison}

The 45 WhAM clips show acoustic gaps from duration-matched natural codas in five of the six encoder representations across alternative donor panels. This comparison includes every available generated clip, but duration matching alone leaves other waveform differences uncontrolled.

\begin{figure}[!htbp]
\centering
\includegraphics[width=\columnwidth]{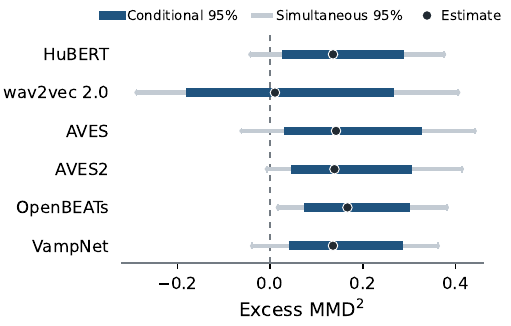}
\caption{Excess MMD$^2$ for 12 WhAM clips versus two disjoint natural panels of 12 codas, matched on waveform properties. Dots show estimates; dark and light bars show conditional and simultaneous 95\% intervals. The dashed line marks zero.\label{fig:mmd}}
\end{figure}

Stronger matching retains 12 generated clips that fall within the natural ranges of the measured waveform properties. We compare them with two disjoint panels of 12 natural codas each. The largest standardized difference is 0.156, below the prespecified limit of 0.20. After common digital processing, the excess acoustic-distribution gaps are positive for all six encoders. Five conditional 95\% intervals exclude zero, while only OpenBEATs excludes zero after simultaneous correction across encoders (Fig.~\ref{fig:mmd}).

The matched comparison is narrower because 33 generated clips lack the required waveform support. Its wider matching caliper and associated checks are reported in the Supplement. Digital processing also cannot reproduce the original recording equipment or geometry. The measured gaps therefore describe these recordings; they do not establish why the generated and natural sounds differ.

\section{Discussion}

\subsection{What the comparison with music reveals}

A sequence of whole units does not reveal how the sounds within each unit are arranged. In codas, click composition adds substantial information beyond timing. Click-form order also contains information against a first-order reference, although the separate held-out order test does not show a corrected gain. Groove combines composition and ordering differently, while the melodic references show smaller or less consistent composition effects. These differences would be missed if the comparison stopped at complete-unit sequences.

The whole-unit analysis asks a different question. Knowing an additional preceding coda improves fitted prediction, while the source-level test remains limited by four conservative source pools. The fitted first-order control finds additional conditional information; the stricter adjacent-pair-preserving reference shows a smaller excess that does not survive the wider correction. These results separate patterns detectable within the recorded sequences from gains established across independent sources. Together with the constituent results, they extend the two-level question in \emph{Rhythm}~\cite{sinha2026} to music without treating coda and musical units as the same.

\subsection{Why this leads to an acoustic blueprint}

Click count and spacing describe when sounds occur, but not the full acoustic form of a coda. Earlier waveform work showed that individual clicks differ beyond timing. Here, composition measures which forms are present, ordering tests how they are arranged, and whole-coda prediction measures relationships across a sequence. Rearranging click placement further changes the encoded sound even when the original click cores and set of intervals are retained. Human-reference directions capture another broad range of recorded variation. Each measurement answers a different acoustic question, so none can stand in for the others.

We bring these measurements together as an \emph{acoustic blueprint} of natural codas. It describes what sounds make up a coda, how those sounds are placed, how whole codas relate across sequences, and which other acoustic differences remain measurable. The musical comparisons show why the parts should be examined separately. The blueprint does not assume that sperm whales share musical categories; it provides a way to ask the same acoustic questions of different recordings.

\subsection{What the synthetic comparison adds}

The WhAM comparison shows how the same measurements can evaluate generated recordings. Duration matching reveals acoustic gaps among the 45 available clips. Stronger waveform matching retains 12 clips, with positive conditional intervals in five encoders and a positive simultaneous interval in OpenBEATs. These differences are measured from the recordings, rather than inferred from whether generated clips sound recognizable.

The tighter comparison also has a clear boundary. Only 12 generated clips have the required measured support, and common digital processing does not recover the original recording geometry or equipment. We cannot infer which part of the generator caused the gap. The result shows what the acoustic comparison detects within the supported recordings, rather than claiming that all generated codas differ for the same reason.

\section{Limitations}

The comparison concerns the 1,952 annotated CETI crops and the three musical references studied here. The four conservative CETI exclusion pools are not four independently identified whales. Dates, callers, social units, clan membership, and published coda-type labels are not established for this cohort. The annotation order does not establish continuous behavioral encounters or same-caller sequences. Native musical event definitions also differ, and the learned states are analysis labels rather than demonstrated biological or musical categories.

State resolution, repetition handling, and the predictor chosen affect the estimates of sequence context. A positive conditional bootstrap interval from four source pools is not an exact source-level significance test. The spectral checks do not remove the uncertainties introduced by recording geometry or acquisition conditions. The human-reference results describe variation along directions validated using human ratings, not the emotions or intentions of whales.

The WhAM comparison concerns fixed isolated clips. Although 12 generated clips can be balanced against disjoint natural donors on measured waveform properties, the original native acquisition conditions remain unknown or nonoverlapping. The matching uses a wider pair caliper than the originally attempted rule, and 33 generated clips lack marginal support. The result cannot identify acquisition-independent synthesis fidelity or the cause of a measured gap. The archive's source hashes, supporting tests, and analysis-specific denominators are documented in the Supplement.

\section{Conclusion}

The comparison with violin music, drum grooves, and sung melodies shows why complete sequences alone do not describe acoustic organization. For codas, the click forms present add information beyond timing. Their order carries conditional information, while the separate held-out order test remains uncertain after correction. Earlier complete codas also improve fitted prediction, with limited source-level evidence. The musical references combine these relationships differently, even though the same questions can be asked of each recording.

The resulting \emph{acoustic blueprint} describes what a coda contains, how its clicks are placed, how complete codas relate across sequences, and how the recorded acoustics vary in other measurable ways. Its parts are complementary rather than interchangeable. The supported WhAM comparison further shows how such measurements can reveal gaps between natural and generated audio without assigning those gaps a cause. Together, the findings extend the study of codas beyond click count and timing while keeping the comparison grounded in recorded sound.

\section*{Supplementary Material}
The accompanying Supplement provides detailed methods, control and sensitivity analyses, data sources and provenance, and additional figures and tables.

\begin{acknowledgments}
The authors acknowledge Project CETI and the researchers who curated and released the recordings, click annotations, and musical reference materials used in this secondary analysis. The companion \emph{Rhythm} work provides scientific context, and the authors thank the teams responsible for the public recordings. No caller or social-unit identities were assigned without supporting metadata.
\end{acknowledgments}

\section*{Author Declarations}

\subsection*{Conflict of Interest}
The authors declare no conflicts of interest related to this work.

\subsection*{Ethics Approval}
This study is a secondary computational analysis of previously collected and released acoustic recordings and annotations. No new animal handling, tagging, playback, or field intervention was performed. Permissions and ethical oversight for the original field recordings remain those of their source projects. This secondary study did not involve new animal contact. Redistribution of restricted waveforms remains subject to the original recording rights and permissions.

\subsection*{Use of AI-Assisted Tools}
AI-assisted language tools were used for editorial restructuring, journal-format conversion, and language refinement. The authors reviewed all scientific content, numerical results, citations, interpretations, and final wording and take full responsibility for the manuscript.

\section*{Data Availability}
The 63 source recordings and their matching annotations are publicly available in the \href{https://drive.google.com/drive/folders/1HkGZcpYbrft3pGtqdt45bUZXVcrJVrdm}{\emph{SpermWhales} collection} linked in reference 34 of Gubnitsky and Diamant~\cite{gubnitsky2024}. These recordings provided the 1,952 codas selected for this study. The Supplement gives the file-level manifest with original filenames, public file identifiers, sizes, and SHA-256 checksums. The companion \emph{Rhythm} analysis uses the same public source collection to study the acoustic organization of clicks and codas~\cite{sinha2026}. The music references include performed Bach violin, Groove~\cite{gillick2019}, and Meertens MTC-FS-1.0 symbolic transcriptions linked to MTC-OGLAUDIO-1.0 field recordings~\cite{vanKranenburg2014}. Human-audio references are DEAM and EmoSoundscapes~\cite{aljanaki2017,fan2017}, and the fixed synthetic clips are from WhAM~\cite{paradise2025}. The accompanying package includes summarized results and figure inputs. The Supplement describes the source checks and links between the reported analyses and the private research archive. The publicly shared source audio and annotations remain with their provider and are not redistributed here. The authors hold the analysis scripts and further derived records privately and will share materials they are permitted to provide on reasonable request. They intend to release these materials publicly after acceptance, subject to third-party rights.

\renewcommand{\theendnotes}{}
\bibliography{references}

@article{aljanaki2017,
  author = {Aljanaki, Anna and Yang, Yi-Hsuan and Soleymani, Mohammad},
  title = {Developing a benchmark for emotional analysis of music},
  journal = {PLOS ONE},
  year = {2017},
  volume = {12},
  number = {3},
  pages = {e0173392},
  doi = {10.1371/journal.pone.0173392}
}

@inproceedings{baevski2020,
  author = {Baevski, Alexei and Zhou, Yuhao and Mohamed, Abdelrahman and Auli, Michael},
  title = {wav2vec 2.0: A framework for self-supervised learning of speech representations},
  booktitle = {Advances in Neural Information Processing Systems},
  year = {2020},
  volume = {33},
  pages = {12449--12460}
}

@inproceedings{fan2017,
  author = {Fan, Jianyu and Thorogood, Miles and Pasquier, Philippe},
  title = {Emo-Soundscapes: A dataset for soundscape emotion recognition},
  booktitle = {Proceedings of the 7th International Conference on Affective Computing and Intelligent Interaction},
  year = {2017},
  pages = {196--201},
  doi = {10.1109/ACII.2017.8273600}
}

@inproceedings{garcia2023,
  author = {Flores Garc{\'i}a, Hugo and Seetharaman, Prem and Kumar, Rithesh and Pardo, Bryan},
  title = {{VampNet}: Music generation via masked acoustic token modeling},
  booktitle = {Proceedings of the 24th International Society for Music Information Retrieval Conference},
  year = {2023},
  pages = {359--366}
}

@article{gero2016,
  author = {Gero, Shane and Whitehead, Hal and Rendell, Luke},
  title = {Individual, unit and vocal clan level identity cues in sperm whale codas},
  journal = {Royal Society Open Science},
  year = {2016},
  volume = {3},
  number = {1},
  pages = {150372},
  doi = {10.1098/rsos.150372}
}

@inproceedings{gillick2019,
  author = {Gillick, Jon and Roberts, Adam and Engel, Jesse and Eck, Douglas and Bamman, David},
  title = {Learning to groove with inverse sequence transformations},
  booktitle = {Proceedings of the 36th International Conference on Machine Learning},
  year = {2019},
  volume = {97},
  series = {Proceedings of Machine Learning Research},
  pages = {2269--2279}
}

@article{gretton2012,
  author = {Gretton, Arthur and Borgwardt, Karsten M. and Rasch, Malte J. and Sch{\"o}lkopf, Bernhard and Smola, Alexander},
  title = {A kernel two-sample test},
  journal = {Journal of Machine Learning Research},
  year = {2012},
  volume = {13},
  pages = {723--773}
}

@inproceedings{hagiwara2023,
  author = {Hagiwara, Masato},
  title = {{AVES}: Animal Vocalization Encoder based on Self-Supervision},
  booktitle = {Proceedings of the IEEE International Conference on Acoustics, Speech and Signal Processing},
  year = {2023},
  pages = {1--5},
  doi = {10.1109/ICASSP49357.2023.10095642}
}

@article{holm1979,
  author = {Holm, Sture},
  title = {A simple sequentially rejective multiple test procedure},
  journal = {Scandinavian Journal of Statistics},
  year = {1979},
  volume = {6},
  pages = {65--70}
}

@article{hsu2021,
  author = {Hsu, Wei-Ning and Bolte, Benjamin and Tsai, Yao-Hung Hubert and Lakhotia, Kushal and Salakhutdinov, Ruslan and Mohamed, Abdelrahman},
  title = {{HuBERT}: Self-supervised speech representation learning by masked prediction of hidden units},
  journal = {IEEE/ACM Transactions on Audio, Speech, and Language Processing},
  year = {2021},
  volume = {29},
  pages = {3451--3460},
  doi = {10.1109/TASLP.2021.3122291}
}

@inproceedings{paradise2025,
  author = {Paradise, Orr and Chen, Liangyuan and Muralikrishnan, Pranav and Flores Garc{\'i}a, Hugo and Pardo, Bryan and Diamant, Roee and Gruber, David F. and Gero, Shane and Goldwasser, Shafi},
  title = {{WhAM}: Towards A Translative Model of Sperm Whale Vocalization},
  booktitle = {Advances in Neural Information Processing Systems},
  year = {2025},
  volume = {38},
  pages = {29269--29305},
  doi = {10.52202/085713-0871}
}

@article{rendell2003,
  author = {Rendell, Luke E. and Whitehead, Hal},
  title = {Vocal clans in sperm whales ({Physeter macrocephalus})},
  journal = {Proceedings of the Royal Society of London B},
  year = {2003},
  volume = {270},
  number = {1512},
  pages = {225--231},
  doi = {10.1098/rspb.2002.2239}
}

@article{sharma2024,
  author = {Sharma, Pratyusha and Gero, Shane and Payne, Roger and Gruber, David F. and Rus, Daniela and Torralba, Antonio and Andreas, Jacob},
  title = {Contextual and combinatorial structure in sperm whale vocalisations},
  journal = {Nature Communications},
  year = {2024},
  volume = {15},
  pages = {3617},
  doi = {10.1038/s41467-024-47221-8}
}

@misc{sinha2026,
  author = {Sinha, Mudit and Chavan, Sanika},
  title = {{Rhythm of the Deep}: Two-tier acoustic organization of sperm-whale codas from click waveforms to coda sequences},
  year = {2026},
  howpublished = {arXiv preprint arXiv:2606.16084v4},
  doi = {10.48550/arXiv.2606.16084}
}

@article{watkins1977,
  author = {Watkins, William A. and Schevill, William E.},
  title = {Sperm whale codas},
  journal = {J. Acoust. Soc. Am.},
  year = {1977},
  volume = {62},
  number = {6},
  pages = {1485--1490},
  doi = {10.1121/1.381678}
}

@article{weilgart1997,
  author = {Weilgart, Linda and Whitehead, Hal},
  title = {Group-specific dialects and geographical variation in coda repertoire in {South Pacific} sperm whales},
  journal = {Behavioral Ecology and Sociobiology},
  year = {1997},
  volume = {40},
  number = {5},
  pages = {277--285},
  doi = {10.1007/s002650050343}
}

@article{mohl2000,
  author = {M{\o}hl, Bertel and Wahlberg, Magnus and Madsen, Peter T. and Miller, Lee A. and Surlykke, Annemarie},
  title = {Sperm whale clicks: Directionality and source level revisited},
  journal = {J. Acoust. Soc. Am.},
  year = {2000},
  volume = {107},
  number = {1},
  pages = {638--648},
  doi = {10.1121/1.428329}
}

@article{zimmertyack2005,
  author = {Zimmer, Walter M. X. and Tyack, Peter L. and Johnson, Mark P. and Madsen, Peter T.},
  title = {Three-dimensional beam pattern of regular sperm whale clicks confirms bent-horn hypothesis},
  journal = {J. Acoust. Soc. Am.},
  year = {2005},
  volume = {117},
  number = {3},
  pages = {1473--1485},
  doi = {10.1121/1.1828501}
}

@article{schulz2009,
  author = {Schulz, Tyler M. and Whitehead, Hal and Rendell, Luke},
  title = {Off-axis effects on the multi-pulse structure of sperm whale coda clicks},
  journal = {J. Acoust. Soc. Am.},
  year = {2009},
  volume = {125},
  number = {3},
  pages = {1768--1773},
  doi = {10.1121/1.3075598}
}

@article{pavan2000,
  author = {Pavan, Gianni and Hayward, Thomas J. and Borsani, Junio Fabrizio and Priano, M. and Manghi, M. and Fossati, Claudio and Gordon, Jonathan},
  title = {Time patterns of sperm whale codas recorded in the {Mediterranean Sea} 1985--1996},
  journal = {J. Acoust. Soc. Am.},
  year = {2000},
  volume = {107},
  number = {6},
  pages = {3487--3495},
  doi = {10.1121/1.429419}
}

@article{antunes2011,
  author = {Antunes, Ricardo and Schulz, Tyler M. and Gero, Shane and Whitehead, Hal and Gordon, Jonathan C. D. and Rendell, Luke E.},
  title = {Individually distinctive acoustic features in sperm whale codas},
  journal = {Animal Behaviour},
  year = {2011},
  volume = {81},
  number = {4},
  pages = {723--730},
  doi = {10.1016/j.anbehav.2010.12.019}
}

@article{oliveira2016,
  author = {Oliveira, Cl{\'a}udia and Wahlberg, Magnus and Silva, M{\'o}nica A. and Johnson, Mark and Antunes, Ricardo and Wisniewska, Danuta M. and Fais, Andrea and Gon{\c c}alves, Jo{\~a}o and Madsen, Peter T.},
  title = {Sperm whale codas may encode individuality as well as clan identity},
  journal = {J. Acoust. Soc. Am.},
  year = {2016},
  volume = {139},
  number = {5},
  pages = {2860--2869},
  doi = {10.1121/1.4949478}
}

@article{begus2025,
  author = {Begu{\v s}, Ga{\v s}per and Sprouse, Ronald L. and Leban, Andrej and Silva, Miles and Gero, Shane},
  title = {Vowel- and diphthong-like spectral patterns in sperm whale codas},
  journal = {Open Mind},
  year = {2025},
  volume = {9},
  pages = {1849--1874},
  doi = {10.1162/OPMI.a.252}
}

@article{friberg1999,
  author = {Friberg, Anders and Sundberg, Johan},
  title = {Does music performance allude to locomotion? {A} model of final ritardandi derived from measurements of stopping runners},
  journal = {J. Acoust. Soc. Am.},
  year = {1999},
  volume = {105},
  number = {3},
  pages = {1469--1484},
  doi = {10.1121/1.426687}
}

@article{mohl2003,
  author = {M{\o}hl, Bertel and Wahlberg, Magnus and Madsen, Peter T. and Heerfordt, Anders and Lund, Anders},
  title = {The monopulsed nature of sperm whale clicks},
  journal = {J. Acoust. Soc. Am.},
  year = {2003},
  volume = {114},
  number = {2},
  pages = {1143--1154},
  doi = {10.1121/1.1586258}
}

@article{thode2002,
  author = {Thode, Aaron and Mellinger, David K. and Stienessen, Sarah and Martinez, Anthony and Mullin, Keith},
  title = {Depth-dependent acoustic features of diving sperm whales ({Physeter macrocephalus}) in the {Gulf of Mexico}},
  journal = {J. Acoust. Soc. Am.},
  year = {2002},
  volume = {112},
  number = {1},
  pages = {308--321},
  doi = {10.1121/1.1482077}
}

@article{rendell2005,
  author = {Rendell, Luke E. and Whitehead, Hal},
  title = {Spatial and temporal variation in sperm whale coda vocalizations: Stable usage and local dialects},
  journal = {Animal Behaviour},
  year = {2005},
  volume = {70},
  number = {1},
  pages = {191--198},
  doi = {10.1016/j.anbehav.2005.03.001}
}

@article{pearce2012,
  author = {Pearce, Marcus T. and Wiggins, Geraint A.},
  title = {Auditory expectation: The information dynamics of music perception and cognition},
  journal = {Topics in Cognitive Science},
  year = {2012},
  volume = {4},
  number = {4},
  pages = {625--652},
  doi = {10.1111/j.1756-8765.2012.01214.x}
}

@techreport{vanKranenburg2014,
  author = {van Kranenburg, Peter and de Bruin, Martine and Grijp, Louis P. and Wiering, Frans},
  title = {The {Meertens Tune Collections}},
  institution = {Meertens Instituut},
  series = {Meertens Online Reports},
  number = {2014-1},
  address = {Amsterdam},
  year = {2014},
  url = {https://www.liederenbank.nl/mtc/}
}

@article{patel2003,
  author = {Patel, Aniruddh D.},
  title = {Language, music, syntax and the brain},
  journal = {Nature Neuroscience},
  year = {2003},
  volume = {6},
  number = {7},
  pages = {674--681},
  doi = {10.1038/nn1082}
}

@inproceedings{openbeats2025,
  author = {Bharadwaj, Shikhar and Cornell, Samuele and Choi, Kwanghee and Fukayama, Satoru and Shim, Hye-jin and Deshmukh, Soham and Watanabe, Shinji},
  title = {{OpenBEATs}: A Fully Open-Source General-Purpose Audio Encoder},
  booktitle = {Proceedings of the IEEE Workshop on Applications of Signal Processing to Audio and Acoustics},
  year = {2025},
  doi = {10.1109/WASPAA66052.2025.11230965}
}

@inproceedings{miron2026,
  author = {Miron, Marius and Robinson, David and Alizadeh, Milad and Gilsenan-McMahon, Ellen and Narula, Gagan and Chemla, Emmanuel and Cusimano, Maddie and Effenberger, Felix and Hagiwara, Masato and Hoffman, Benjamin and Keen, Sara and Kim, Diane and Lawton, Jane and Liu, Jen-Yu and Raskin, Aza and Pietquin, Olivier and Geist, Matthieu},
  title = {{AVEX}: What Matters for Animal Vocalization Encoding},
  booktitle = {Proceedings of the International Conference on Learning Representations},
  year = {2026},
  url = {https://openreview.net/forum?id=MFuM9KAEYc}
}

@article{repp1992,
  author = {Repp, Bruno H.},
  title = {Diversity and commonality in music performance: An analysis of timing microstructure in {Schumann}'s {Tr{\"a}umerei}},
  journal = {J. Acoust. Soc. Am.},
  year = {1992},
  volume = {92},
  number = {5},
  pages = {2546--2568},
  doi = {10.1121/1.404425}
}

@article{russell1980,
  author = {Russell, James A.},
  title = {A circumplex model of affect},
  journal = {Journal of Personality and Social Psychology},
  year = {1980},
  volume = {39},
  number = {6},
  pages = {1161--1178},
  doi = {10.1037/h0077714}
}

@article{juslin2003,
  author = {Juslin, Patrik N. and Laukka, Petri},
  title = {Communication of emotions in vocal expression and music performance: Different channels, same code?},
  journal = {Psychological Bulletin},
  year = {2003},
  volume = {129},
  number = {5},
  pages = {770--814},
  doi = {10.1037/0033-2909.129.5.770}
}

@article{gubnitsky2025,
  author = {Gubnitsky, Guy and Mevorach, Yaly and Gero, Shane and Gruber, David F. and Diamant, Roee},
  title = {Automatic detection and annotation of eastern {C}aribbean sperm whale codas},
  journal = {Sci. Rep.}, volume = {15}, number = {1}, pages = {12790}, year = {2025},
  doi = {10.1038/s41598-025-97009-z}
}

@misc{gubnitsky2024,
  author = {Gubnitsky, Guy and Diamant, Roee},
  title = {Detecting the Presence of Sperm Whales' Echolocation Clicks in Noisy Environments},
  year = {2024},
  howpublished = {arXiv preprint arXiv:2401.00900v1},
  url = {https://arxiv.org/abs/2401.00900v1},
  note = {Reference 34 links the publicly shared SpermWhales recording and annotation collection}
}

\end{document}